\documentclass[checkin,showpacs,aps,prl,showkeys]{revtex4-1}
\usepackage{threeparttable}
\usepackage{amsmath}
\usepackage{lipsum}
\usepackage{multirow}
\usepackage{booktabs}
\usepackage{graphicx,color}
\usepackage{graphicx}
\usepackage{subfigure}
\usepackage{amsmath}

\newcommand{\ba}{\begin{eqnarray}}
\newcommand{\ea}{\end{eqnarray}}

\newcommand{\be}{\begin{equation}}
\newcommand{\ee}{\end{equation}}

\definecolor{pink}{rgb}{1,0.18,1.0}
\def\prl{{ Phys. Rev. Lett. }}

\def\apl{{ Appl. Phys. Lett. }}
\def\prb{{ Phys. Rev. B }}

\def\njp{{N. J. Phys. }}

\def\jpcm{{J. Phys.: Condens. Matter }}

\def\jpcc{{J. Phys. Chem. C }}

\def\pr{{Phys. Rev. }}
\def\nl{{Nano Lett. }}
\def\nn{{Nat. Nanotech. }}
\def\acsnn{{ACS Nano }}

\def\nc{{Nat. Commun. }}

\def\jpcc{{J. Phys. Chem. C }}
\def\jpcl{{J. Phys. Chem. Lett. }}

\def\2dm{{2D Materials}}
\def\sr{{Sci. Rep. }}
\def\apex{{Appl. Phys. Express }}
\def\am{{Advanced Materials}}
\def\afm{{Adv. Funct. Mater}}

\begin{document}

\title{A large enhancement of carrier mobility in phosphorene by
  introducing hexagonal boron nitride substrate}

\author{Jiafeng Xie}
\affiliation{Key Laboratory for Magnetism and Magnetic Materials of
 the Ministry of Education, Lanzhou University, Lanzhou 730000, China}

\author{Z. Y. Zhang}
\affiliation{Key Laboratory for Magnetism and Magnetic Materials of
 the Ministry of Education, Lanzhou University, Lanzhou 730000, China}

\author{D. Z. Yang}
\affiliation{Key Laboratory for Magnetism and Magnetic Materials of
 the Ministry of Education, Lanzhou University, Lanzhou 730000, China}

\author{M. S. Si$^{*}$}
\affiliation{Key Laboratory for Magnetism and Magnetic Materials of
 the Ministry of Education, Lanzhou University, Lanzhou 730000, China}

\author{D. S. Xue}
\affiliation{Key Laboratory for Magnetism and Magnetic Materials of
 the Ministry of Education, Lanzhou University, Lanzhou 730000, China}

\author{Xiaohui Deng}
\affiliation{Department of Physics and Electronic Information Science,
  Hengyang Normal University}

\date{\today}

\begin{abstract}
Carrier mobility is a crucial character for electronic devices
since it domains power dissipation and switching speed.
Materials with certain high carrier mobility, equally,
unveil rich unusual physical phenomena
elusive in their conventional counterparts.
As a consequence, the methods to enhance
the carrier mobility of materials receive
immense research interests due to
their potential applications in more effective electronic devices
and enrichment of more unusual phenomena.
For instance, introducing a flat hexagonal boron nitride (h-BN)
substrate to enhance
the carrier mobility has been achieved experimentally.
However, the underlying mechanics is not well understood.
 In this study, we estimate the carrier
mobility of phosphorene on h-BN substrate (P/h-BN) within the framework of
the phonon-limited scattering model at first-principles level.
Besides high-$\kappa$ dielectric property, h-BN also
possesses excellent mechanical property of a high two-dimensional elastic modulus.
The P/h-BN heterostructure inherits the high elastic modulus of h-BN, leading
to an enhanced carrier mobility in phosphorene. Owing to the weak
van der Waals interactions between the layers, the unique electronic
properties of phosphorene are almost perfectly preserved near the Fermi level,
guaranteeing the superior electronic transport in P/h-BN. Our findings
offer a new perspective to improve the carrier
mobility in phosphorene as well as other 2D materials based field effect transistors.
\end{abstract}


\maketitle

\section{I. Introduction}

Phosphorene, isolated from black phosphorus,
exhibits many tantalizing properties, which are not
present in the bulk counterpart \cite{ling}.
As a promising semiconductor,
phosphorene exhibits a considerable direct band gap, which is in contrast
to gapless graphene \cite{liu}. More importantly,
compared to monolayer of transition metal
dichalcogenides with direct band gaps, carrier mobility of phosphorene
behaves in a more superior manner \cite{radi,qiao}.
Carrier mobility
is an important parameter for semiconducting materials as it
dominates the power dissipation and switching speed
\cite{dasgu}.
As a consequence, the phosphorene based device may offer a new
opportunities to promote the desire of electronics and optoelectronics
in the post-silicon era \cite{wang}.
Further move,
many people in the community are looking different
aspect on phosphorene
to the enhancement of
carrier mobility in phosphorene.
For instance, introducing a flat hexagonal boron nitride
(h-BN) substrate to confine the
two-dimensional (2D) electron gas \cite{dean}.

Phosphorene field effect transistors (FETs) were first
synthesized on a SiO$_{2}$/Si substrate. The hole mobility
of phosphorene reaches 286 cm$^{2}$V$^{-1}$s$^{-1}$
in $\sim$4 nm film at room temperature,
while it decreases to
24 cm$^{2}$V$^{-1}$s$^{-1}$ in $\sim$1 nm film
\cite{liu}.
By introducing the h-BN substrate,
the hole mobility in $\sim$1 nm phosphorene film
rises up to 400 cm$^{2}$V$^{-1}$s$^{-1}$
 \cite{li}.
That means the h-BN substrate induce near 20 times
enhancement of hole mobility
for the actual phosphorene based device applications.
In addition, rich unusual
physical phenomena, such as quantum oscillations and quantum Hall
effect \cite{long}, can only be probed in phosphorene FETs under a
certain high  carrier mobility.
However, the mechanism
of the enhanced
carrier mobility is not well understood.
Some groups owe the higher carrier mobilities in phosphorene FETs
to fewer layers and lower temperatures
taken in their experiments \cite{li_14,xia}.
In this letter, we ignor the effects of layers and temperatures,
but pay more attention to intrinsic mechanics.

Theoretical explanations on enhancement of carrier mobility
are divergent.
In 2007,
Jena and Konar
 demonstrated that the carrier mobility
can be enhanced by modifying the dielectric environment
\cite{jena}.
 The high-$\kappa$ dielectric materials would increase the
damping of Coulombic scattering, resulting in a higher carrier
mobility.
However,
Ponomarnenko and colleagues experimentally observed no obvious
enhancement in carrier mobility of graphene on different dielectric
substrates in 2009 \cite{pono}.
That means the explanation of high-$\kappa$ dielectric environment may
not suitable for graphene.
Soon,
Konar and coauthors propose a new theory
that,
owing to the confined 2D electron gas,
graphene holds a strong surface phonon
scattering, washing out the partial screening of Coulombic scattering
\cite{konar}.
In this perspective, the enhanced carrier mobility
cannot be simply ascribed into a single model for all 2D layered
materials. It naturally raises the crucial  question that why the
h-BN substrate can enhance the carrier mobility in phosphorene.

To resolve the puzzle, we construct the heterostructure of
phosphorene/h-BN (P/h-BN) within the framework of the phonon-limited
scattering model at first-principles level.
The obtained results are directly compared
with those of phosphorene. It is found that the enhanced carrier
mobility arises from the increased 2D elastic modulus, which are
inherited from the h-BN substrate. At the same time, the unique
electronic properties of phosphorene are well preserved in P/h-BN due
to the weak van der Waals (vdW) interactions between phosphorene and
h-BN. Our finding allows the control of enhancing carrier mobility
in phosphorene field effect transistors, which is essential in the
design of new devices based on different dielectric substrates.

\section{II. METHODS}

Our first-principles calculations are performed through using the {\sf
  siesta} package within the framework of density functional theory
(DFT) \cite{siesta}. To study phosphorene on h-BN substrate, we use
the generalized gradient approximation for the exchange and
correlation potential as proposed by
Perdew-Burk-Ernzerhof (PBE)
\cite{pbe}. As the vdW interactions are important in predicting the
stable stacking of 2D layered materials, the optB88-vdW functional is
employed in the  realistic calculations \cite{optvdw}. For comparison,
three other functionals of LDA \cite{ca}, vdW-DF2 \cite{df2}, and C09
\cite{c09} are taken to estimate the interlayer binding energy of
P/h-BN heterostructure. The plane wave energy cutoff is set to be 210 Ry to
 ensure the convergence of total energy. The reciprocal space is
 sampled by a fine grid of 11$\times$5$\times$1 {\it k}-point in the
 Brillouin zone. Structural relaxations were performed until the force
 on each individual atom is less than 0.01 eV/{\AA}. The optimized
 double-$\zeta$ orbitals including polarization orbitals are employed
 to describe the valence electrons.

\section{III. Formalism}

To theoretically estimate the carrier mobility of 2D layered materials
is not an easy task as it always involves many scattering processes,
such as the charged-impurity scattering, the phonon scattering as well
as the Coulombic scattering \cite{jena,isberg,hwang}.
For simplicity, we take the so-called phonon-limited
scattering model to calculate the carrier mobility of the 2D
heterostructure of P/h-BN, which reads \cite{takagi,bruzzone,zhiya,pholimit}
\be
\mu_{2D} = \frac{e\hbar^{3}C_{2D}}{k_{B}Tm^{*}m^{*}_{a}E_{i}^{2}},
\ee
where $e$ is the electron charge, $\hbar$ the reduced Plank's
constant, $k_{B}$  the Boltzmann constant and $T$ the temperature,
$m^{*}$ ($m^{*}_{a}$) the (averaged) effective mass with respect to
the transport direction.
$E_{i}$ the deformation potential constant
of valence band maximum (VBM) for hole or conduction band minimum
(CBM) for electron along the transport direction. The latter two
quantities are closely related to band structures. $m^{*}$ is referred
as the second derivative of VBM or CBM band relative to crystal
momentum and the averaged effective mass is determined by
$ m_{a}^{*} = \sqrt{ m^{*}_{x}m^{*}_{y} }$.
Whereas, $E_{i}$ is decided by the shift of VBM or CBM band in response to
the strain applied along the transport direction. In this regard, we
expect phosphorene dominates  these two quantities as phosphorene
holds a sizable direct band gap and other excellent intrinsic
electronic properties including
 effective mass and deformation potential.
$C_{2D}$ is the 2D elastic modulus
 of the longitudinal strain in the propagation directions
 of the longitudinal acoustic (LA) wave.
 It is a purely
mechanical property. As a high-$\kappa$ dielectric material, h-BN can
modify the dielectric environment. Moreover, the  flat h-BN possesses
a higher  $C_{2D}$ and may increase the carrier mobility, as expected
in Eq. (1), if the P/h-BN heterostructure inherits  the $C_{2D}$ of h-BN.

\section{IV. Electronic properties of P/\MakeLowercase{h}-BN}

Considering that the lattice parameters of
phosphorene and h-BN are different,  we
construct the heterostructure of P/h-BN by extending h-BN
 with 1$\times$4 unit cells to match the fixed
phosphorene with  1$\times$3 unit cells during the geometric
optimization, as displayed in Fig. 1(a). In this way,
 the obtained unit cell of P/h-BN
heterostructure has a minimum initial mismatch of $\sim$3\% \cite{fei,cai2}.
It is well known that the stacking manner plays a crucial role in electronic
properties of heterostructure. For diminishing this uncertainty in our
calculations, we firstly search various stackings of P/h-BN for the most
stable one as our starting point. Based on the most stable stacking,
we respectively shift phosphorene relative to h-BN along the
armchair (x) and zigzag (y)
directions. The corresponding energy curves are plotted in
Fig. 1(b).
As can be seen,
when the shift is along the x
direction, it needs to  overcome
a large energy barrier as the puckered
structure of phosphorene is involved (see the bottom panel of
Fig. 1(a)). By contrast, the energy barrier  related to the
y-direction shift is quite small. This is because the flat h-BN
dominates the energy curve and no obvious change in configuration
occurs during this shift process. We also notice that the maximum
energy barrier appears at the shift of around 5\%, which was not
observed in P/graphene heterostructure \cite{padilha}. At this
position, we find that the P atoms of phosphorene are  very close to
the N atoms of h-BN. Since the residual charges of h-BN largely reside
in the N atoms, the Coulomb interactions between the layers are
increased, giving rise to a maximum energy barrier.

In order to accurately estimate the interlayer binding of P/h-BN
heterostructure, we take five exchange-correlation functionals  in our
simulations, as shown in Fig. 1(c). Without the vdW correlations,
the functionals of PBE and LDA cannot assess rightly the
interlayer coupling. The PBE underbinds and increases the interlayer
spacing, while the LDA overbinds and reduces the interlayer
spacing. Even if the vdW correlation is taken into account, the
results are still not converged. The vdW-DF2 method gives  an
interlayer spacing at least larger than 3.7 {\AA}, leading to a
reduced binding energy of around 35 meV. The advanced version of vdW
correlation, that is optB88-vdW, seems to be suited for the case of
P/h-BN heterostructure. The obtained interlayer spacing is around 3.5
{\AA} and the binding energy is around 44 meV, which are in good
agreement with previous calculations \cite{rivero,cai}. Based on the
same functional of optB88-vdW, the binding energies of bilayer
phosphorene and bilayer h-BN are  $\sim$55 meV \cite{cai_sr,dai} and
$\sim$63 meV \cite{hsing}, respectively. This means the interlayer vdW
interaction between phosphorene and h-BN is weaker than those in both
bilayer phosphorene and  bilayer h-BN. Such a weak coupling would
preserve the electronic properties of phosphorene near the Fermi level in P/h-BN
heterostructure. It is also found that this interlayer coupling is
comparable to that of P/graphene \cite{padilha,cai}. In addition, the
vdW functional of C09 produces very similar results
to those of optB88-vdW.  It is
reasonably believed that the optB88-vdW functional can evaluate the
interlayer coupling of the heterostructure of P/h-BN
at first principle calculation level. And it is thus used
for the next calculations. Certainly, more accurate results can be
obtained by  quantum Monte Carlo method \cite{shulenburger}, but the
computational cost is beyond our tolerance.

Next, we check whether the P/h-BN heterostructure did inherit the
electronic properties of phosphorene. For comparison, the band
structure of phosphorene is calculated, as shown in Fig. 2(a). It has
a direct band gap of around 0.9 eV at the $\Gamma$ point. The bands of
CBM and VBM have a stronger dependence on the crystal momentum along
the $\Gamma$-X (armchair) direction  than that
along the $\Gamma$-Y (zigzag)
direction.
All the electronic properties mentioned above
are well in line with previous studies \cite{liu,radi,qiao}.
When phosphorene is placed on h-BN substrate,
the calculated band structure  is displayed in Fig. 2(b). At  first
glance, the bands near the Fermi level are almost the same as those of
phosphorene, which is also  verified by
the PDOS, as shown in Fig. 2(c). In the energy range from -1.5 eV to
2.0 eV, the P-PDOS does not overlap with the B- and N-PDOS, indicating a
very weak coupling between phosphorene and h-BN. A closer look at
these two band structures reveals some slight differences. For
example, more bands of CBM are degenerate at the $\Gamma$ point in
comparison with phosphorene, resulting from the stacking effect of
phosphorene on h-BN substrate. This means  that the CBM bands are very
sensitive to the external perturbation, in line with previous theory
\cite{rodin}. In P/h-BN heterostructure, since more bands are
introduced by the h-BN substrate, the interactions between the bands
are enhanced. Such band interactions push the VBM bands of phosphorene
closer to each other and make them highly degenerate at the $\Gamma$
point. {As stated by Herring \cite{herring}, this degeneracy is
  accidental because bands could degenerate at some specially
  crystal points perturbed by external stimulation. We also noted that
  the degeneracy reported by Hu and coauthors \cite{hu} is smaller
  than that of our model. This is because the structure of phosphorene
  is relaxed during the geometry optimization in Hu's work, which is
  in contrast to the fixed one in our case.}
Another effect from the h-BN substrate is the increased band
gap of P/h-BN. Wider band-gap h-BN introduces an
extra energy of around 0.1 eV into the band gap of P/h-BN
heterostructure. Except for the above mentioned two slight
differences, the P/h-BN heterostructure almost perfectly
inherits the
electronic properties of phosphorene  near the Fermi level as we expected.

According to the effective-mass theorem \cite{ashcroft}, the effective
mass  is associated with the band structure via
$1/m_{n}^{*}= (\partial^{2}/\hbar^{2}\partial k^{2})V_{n}(k)$ with
$V_{n}(k)$ being the band energy of the $n$th
band  at crystal momentum $k$. This
means that  the effective mass of carriers largely depends on the band
curvatures of VBM or CBM. Since the bands near the Fermi level of
P/h-BN heterostructure are almost the same as those of  phosphorene, their
effective masses should be  similar. As illustrated in Table
I, the introduced h-BN substrate  only has an effect on the effective
masses along the $\Gamma$-Y direction,  arising from the enhanced
band interactions along this direction. The effective mass for
electron decreases from 1.12 $m_{0}$ in phosphorene to 0.43 $m_{0}$ in
P/h-BN along the $\Gamma$-Y direction.  The similar trend is also
observed in the hole carrier along this direction. By contrast, the
effective masses along the $\Gamma$-X direction are nearly unchanged,
yielding a reduced transport anisotropy compared to phosphorene.

Since the mobility $\mu$ is inversely
proportional to the square of deformation potential $E_i^2$, the deformation potential
is another key factor to act on the carrier mobility, which
is in connection with the band structure
through $E_{i}=\Delta V_{i}/(\Delta l/l_{0})$ with
$V_{i}$ being the energy change of the $i$th band
under cell
compression or dilatation of $\Delta l$ and $l_{0}$ the lattice
constant.
According to our calculations, though some deformation potentials are
influenced by the stacking of h-BN layer, most of the
deformation potentials for both electron and hole in P/h-BN are
unchanged compared to those of phosphorene (see Table I for more
details). Within the framwork of longitudinal acoustic phonon
perturbation, the dielectric function $\epsilon(\omega,q)$ is
connected to the deformation potential $E_{i}$ via $\Delta
E_{i}=E_{i}[\epsilon^{-1}(\omega,q)-1]$.  According to the analysis
suggested by Pollak and coauthors \cite{pollak}, the main contribution
to the $\epsilon(\omega,q)$ is the transitions from the VBM
bands to the CBM bands.  Thus, the introduced h-BN substrate cannot
change the dielectric environment of phosphorene. This is because
phosphorene dominates the VBM and CBM bands of P/h-BN, leaving the
deformation  potentials nearly unchanged. In this perspective, we can
easily exclude the enhanced carrier mobility in the P/h-BN
heterostructure from the increased damping of Coulombic scattering
\cite{jena}.
Unavoidable,
two deformation potentials are not unchanged,
 the deformation potential for electron along x
direction is increased $\sim$0.8 eV (more than 50\%),
while that for hole along y direction is decreased
$\sim$0.3 eV (nearly 60\% ).
According to the Eq. (1), small deformation potential
leads to high carrier mobility. Thus,
the electron velocity 
is suppressed by the h-BN layer,
while the hole velocity is accelerated.
However, the realization is not that straightforward.
As shown in Tab. \ref{tab1},
the electron mobility is higher than that of hole
along x direction.
Whereas, along y direction, electron mobility loose this advantage.
This is because the carrier mobility is not only sensitive to the
electronic properties, but also related to the mechanics.
In the following, we will demonstrate the power of h-BN
substrate on the enhancement of carrier mobility  from another
viewpoint.

\section{V. 2D elastic moduli of P/\MakeLowercase{h}-BN}

As a high-$\kappa$ dielectric material, h-BN enters current researches
with a main focus on its quantum confinement. For example, other
not-flat layered materials  can be confined on flat h-BN substrate
through vdW stacking, usually
 appear promising properties such as high carrier mobility
\cite{vdw_hetero,bachhuber} .
By contrast, it is always neglected that
its mechanical property
also contribute to the `high mobility'.
In fact, the elastic property of h-BN
has a big effect on the charge transport of vdW
heterostructures. This is really the case of  P/h-BN heterostructure.
As shown in Table I, once the substrate of h-BN is introduced, the 2D
elastic moduli of P/h-BN are significantly improved: $C_{x-2D}$
increases from $\sim$30 Jm$^{-2}$ to $\sim$320 Jm$^{-2}$ and
$C_{y-2D}$ from $\sim$96 Jm$^{-2}$ to $\sim$370
Jm$^{-2}$. This directly induces the enhancement of carrier
mobility according to Eq. (1).
For example, the electron moblity along the x direction is
about five times of that of phosphorene.
It is also noticed that the
  carrier mobilities of electron are smaller than those of hole along
  the y direction, which are in line with the experimental results
  \cite{li}.
These tell us that the enhanced
carrier mobility in the P/h-BN heterostructure is realized via the
increased $C_{2D}$. Why does it happen? We next  reveal  the
underlying mechanism by comparing the 2D elastic moduli of the three
materials, i.e.,  P/h-BN, phosphorene and h-BN.

The 2D elastic modulus, i.e., $C_{2D}$, can be obtained through
fitting the expression  $(E-E_{0})/S_{0}=C_{2D}(\Delta
l/l_{0})^{2}/2$, where $E$ is the total energy and $S_{0}$ is the
equilibrium lattice volume
(the xy surface area)
for a 2D material. First, we estimate the
$C_{2D}$ of phosphorene, as shown in Fig. 3. The strain energy density
curve for the x-direction strain is obviously lower than that for the
y-direction strain, resulting from the structural anisotropy of
phosphorene. The 2D elastic moduli are  $C_{x-2D}$ = 30.24
Jm$^{-2}$ and   $C_{y-2D}$ = 96.16 Jm$^{-2}$, in good agreement with
previously reported values \cite{qiao}. Then, we study the 2D elastic
moduli of P/h-BN heterostructure. As can be seen in Fig. 3, the strain
energy  density curves for P/h-BN heterostructure are largely promoted
in comparison with phosphorene.  Under tension in the range of 0\% to
2\%, the curve of x-direction is slightly different from that of
y-direction. By contrast, this difference disappears under
compression. This means that the P/h-BN heterostructre partially
preserve the structural anisotropy of phosphorene. Such transport
anisotropy only under tension has  not been reported in previous
literatures.  By fitting these curves, two slightly different 2D
elastic moduli of $C_{x-2D}$ = 321.41 Jm$^{-2}$ and  $C_{y-2D}$ =
369.86 Jm$^{-2}$ are easily obtained, which are larger than those of
phosphorene, but are comparable to those of h-BN.
For h-BN, two identical 2D elastic
moduli  of $C_{2D}$ = 292.34 Jm$^{-2}$ are obtained, in
accordance with experimental and other theoretical reports
\cite{song,lee,kudin}. Actually, as shown in Fig. 3, the curves of P/h-BN are
roughly the superposition of h-BN and phosphorene, in which h-BN
dominates. This means that the P/h-BN heterostructure  inherits well
the elastic property, that is the 2D elastic modulus, of h-BN,
inducing the enhanced carrier mobility in phosphorene. According to
our results, further improving the carrier mobility requires
a flat substrate which is a 2D layered material
itself with higher 2D modulus. Future research can test this
prediction.

In pure materials, interactions between the carriers and
the acoustical vibrations dominate the mobilities of
electrons and holes.  The carriers interact
only with those acoustical vibrational modes of long wave-length
whose properties are determined by the elastic modulus
\cite{bardeen,shockley}. The acoustical sound velocity $s_{l}$ in long
wave-length limit is connected to the elastic modulus $C$ via $s_{l} =
C/\rho^{1/2} =(\partial /\partial k)\omega(k)$ with $\rho$ being the
mass density and $\omega(k)$ the phonon frequency. By stacking
phosphorene on h-BN substrate,
the 2D elastic moduli are increased by inheriting the
elastic property of h-BN, directly increasing the acoustical sound velocity. The
corresponding phonon  wave-length of P/h-BN is thus decreased, reducing the
effective electron-phonon scattering. As a result, the carrier
mobilities of P/h-BN are enhanced. This is not surprising. When a soft
material is placed on a hard substrate, the lattice vibration of such
stacked heterostructure is determined by the hard substrate and
is thus suppressed. This demonstrates the underlying mechanism of
the enhanced carrier mobility in P/h-BN, which is in contrast to the theory
laying emphasis on the increased damping of Coulombic scattering
through  modifying the dielectric environments \cite{jena}. At the
same time, we also found that the 2D elastic modulus of SiO$_{2}$ is
smaller than those of h-BN. Based on the above discussion, we can
infer that the carrier mobilities of phosphorene on h-BN are larger
than those on SiO$_{2}$.

It is known that the h-BN, a substrate, in experiment must
be few-layer or even thin film. Thus,
we construct a supercell with monolayer phosphorene on top of bilayer
h-BN (P/2h-BN) heterostructure.
The relative factors are shown in Tab. \ref{tab1}.
As it can be seen, the anisotropy inherit from phosphorene becomes
weaker, both the electronic and the elastic properties,
owing to the isotropic h-BN substrate.
However, with the vdW stacking,
the carrier mobilities of P/2h-BN heterostructure
are not largly improved in both x and y directions as expected.
Especially along y direction, the carriers decrease largely compare with
P/h-BN but still larger than the monolayer phosphorene.
It can be understand that
the h-BN layers induce a larger elastic modulus as well as
a larger deformation potential.
The deformation potential,
which is the band energy change during the strain being applied,
also play
a key roll in carrier mobility calculation.
However, comparing with the effect of deformation potential
induced by h-BN, the large elastic modulus play a
more important role in the carrier mobility of P/2h-BN.
The predicted carrier mobilities of bilayer phosphorene on top
of monolayer h-BN (2P/h-BN) heterostruture in Tab. \ref{tab1}
also state the similar facts that elastic modulus indeed
play a more critical role.

{ Finally, we should emphasize two points. One is referred to the 2D
elastic modulus, which does not increase monotonously with the
thickness of substrate. This is because the 2D elastic modulus is
indeed related to the deformation potential.
Generally speaking,
{ the deformation potential is the energy change of the VBM and CBM during strains being applied, it}
increases with the vdW stacking, which will suppress the carrier mobility.
Another is referred to the predicted carrier mobilities in our work, which are
much larger than those reported experimentally. This is mainly caused
by the introduction of impurity or/and vacancy during the synthesis
process of phosphorene in experiment. We believe
the carrier mobility will be largely improved if phosphorene is prepared
with high quality \cite{nov}.}

\section{VI. Model check}
Since the phonon-limited scattering model has
its validity limits \cite{walu},
we calculated phonon spectra of monolayer phosphorene,
 h-BN, and P/h-BN heterostructure
to further validate the accuracy of our methods,
as shown in Fig. \ref{fig4}.
For more comparable, the supercell
are used in calculations.
With 12 atoms, phosphorene has 36 vibration branches
 in its phonon spectra (Fig. \ref{fig4}(a)).
The 3 acoustic branches:
inplane transverse acoustic (TA, along x direction), the LA (along y direction)
mode, and the out-of-plane acoustic (ZA) mode, show linear rise near
the $\Gamma$ point.
And a gap of phosphorene vibration (stop band) appears near $\omega$=300 cm$^{-1}$,
which are in good agreement with
previous studies \cite{zhu,jain}.
Comparing with the phonon vibrations of phosphorene,
those of h-BN exhibit higher vibration frequencies (both acoustic modes and optical
modes) \cite{michel,jung},
as shown in Fig. \ref{fig4}(b).
This is because the B-N bonds are
shorter in length and stronger in polarity than P-P bonds.
The larger slopes of the LA branches near $\Gamma$ of h-BN indicate
the higher speed of sound and the in-plane stiffness.
Also, the stop band of h-BN is smaller than that of
phosphorene, indicating that h-BN have larger thermal
conductivity \cite{mcgau}. 
As for P/h-BN heterostructure,
it exhibits more vibration modes and smaller stop band comparing with phosphorene,
which inherit from h-BN,
as shown in Fig. \ref{fig4}(c).
That means larger thermal conductivity 
will also be inherited.
Moreover, the acoustic modes of P/h-BN heterostructure
are suppression comparing with phosphorene,
resulting in larger in-plane stiffness and larger elastic modulus.
This can be understand that
the out-of-plane vibration modes of phosphorene in P/h-BN are
suppressed by the flat h-BN,
which reduce the slop of acoustic modes and stop band in heterostructure,
leading to a larger carrier mobility.

\section*{Conclusions}

In conclusion, we have performed first-principles calculations on the
P/h-BN heterostructure to reveal the mechanism of the enhaced carrier
mobility in phosphorene on h-BN substrate. The enhanced carrier
mobility is demonstrated to be correlated with  the increased 2D
elastic moduli inherited  from the h-BN substrate. The unique
electronic properties of phosphorene are almost preserved in the
P/h-BN heterostructure, which is due to the weak vdW interactions
between phosphorene and h-BN. Our results inevitably offer a new
perspective to improve the carrier mobilities  in phosphorene field
effect transistors.

\section*{ACKNOWLEDGMENTS}

We would like to thank Zhen Zhu (MSU)
for valuable communications.
This work was supported by the National Basic Research Program of
China under Grant No. 2012CB933101 and  the National Science
Foundation under Grant No. 51372107,
 No. 11104122 and No. 51202099.
This work was also supported by the National Science Foundation for
Fostering Talents in Basic Research of the National Natural Science
Foundation of China.

$^{*}$sims@lzu.edu.cn

\clearpage

\begin{figure}
\includegraphics[width=15cm]{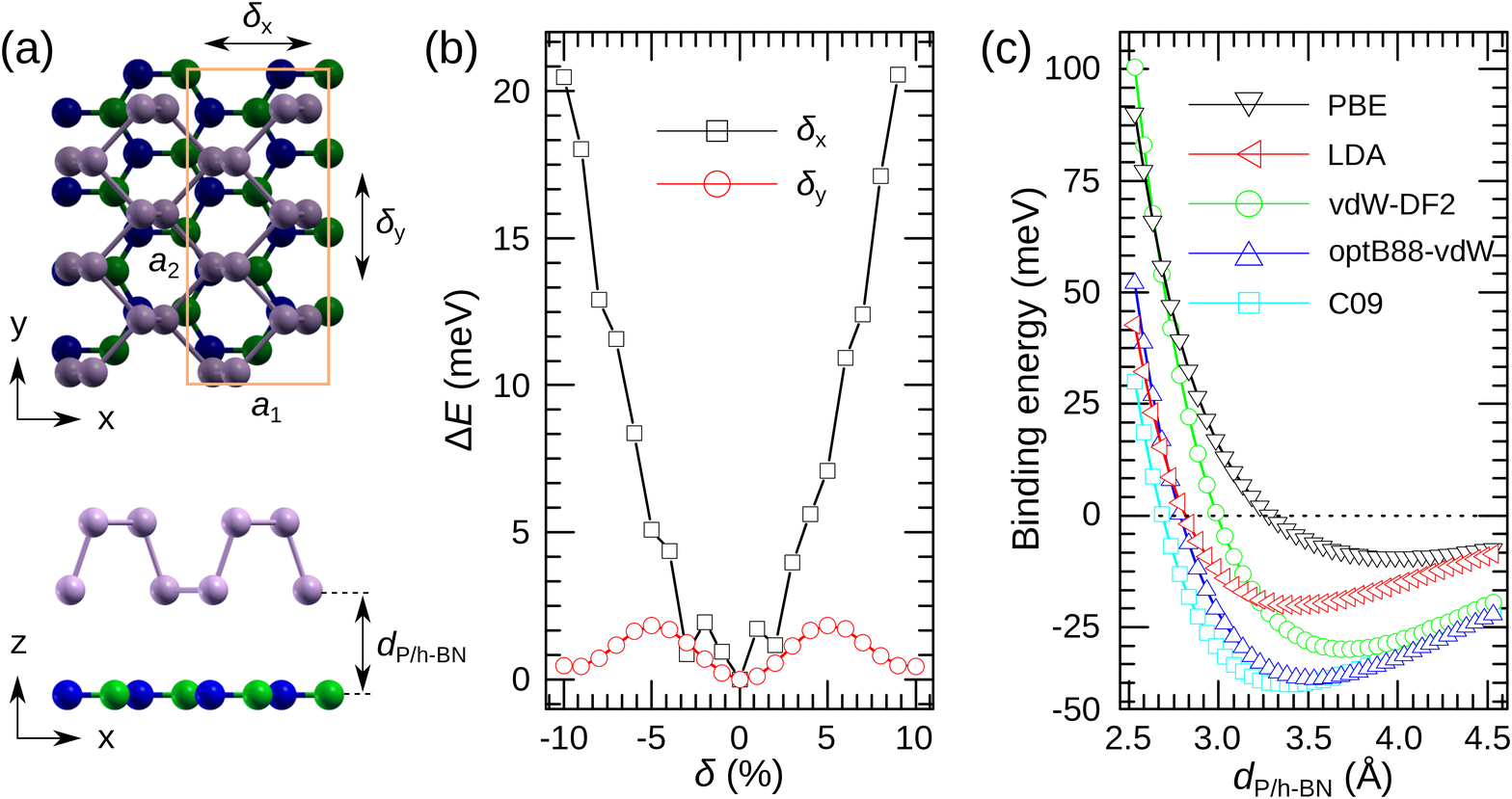}
\caption{(color online). (a) Top (top panel) and side (bottom panel) views of P/h-BN heterostructure.
 The wihte, blue and green balls represent phosphorus, boron and nitrogen atoms, respectively.
 The yellow rectangle represents the unit cell {\color{black} with $a_{1}$ = 4.54 {\AA} and $a_{2}$ = 10.04 {\AA}}.
 $\delta_{x}$ and  $\delta_{y}$ are the
 relative displacements of phosphorene to h-BN. $d_{\rm P/h-BN}$ is defined as the distance
 between the phosphorene and h-BN layers. (b) Total energy change as a function of
 displacements $\delta_{x}$ and $\delta_{y}$.  The functional of optB88-vdW is taken
 in this calculation. (c) Binding energy per atom as a function of the
 interlayer spacing  $d_{\rm P/h-BN}$ for five exchange-correlation functionals.}
 \label{fig1}
\end{figure}

\clearpage

\begin{figure}
\includegraphics[width=16cm]{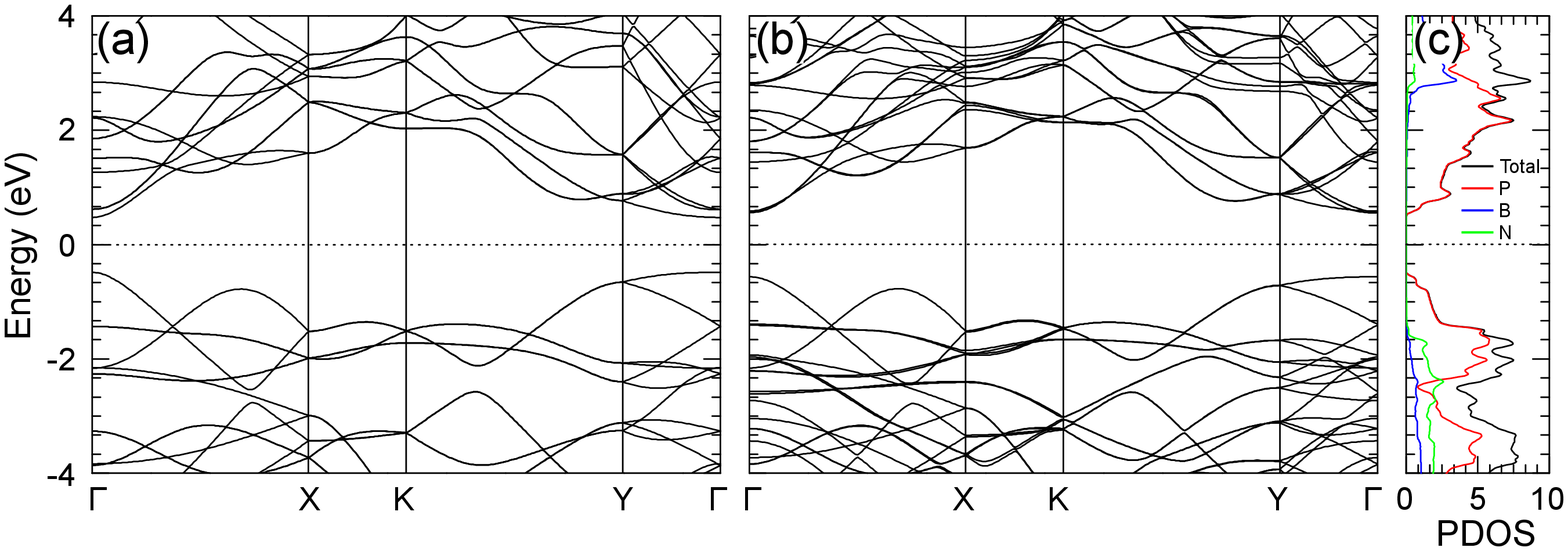}
\caption{(color online). Band structures of (a) phosphorene  and (b) P/h-BN heterostructure.
 (c) Projected density of states (PDOS) of P/h-BN,
 where the black, red, blue, and green lines represent the total density of states,
 P-, B-, and N-PDOS, respectively. The Fermi level is set to energy zero. }
 \label{fig2}
\end{figure}

\clearpage

\begin{figure}
\includegraphics[width=10cm]{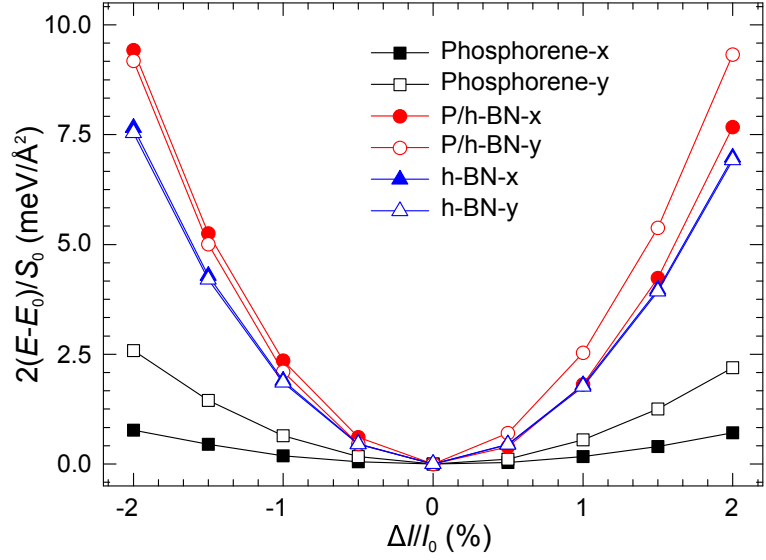}
\caption{(color online). The 2D elastic moduli are obtained by fitting
  the strain energy density curves $2(E-E_{0})/S_{0}$ versus $\Delta
  l/l_{0}$ for phosphorene, P/h-BN and h-BN. The strain is applied
  along the x and y directions.}
\label{fig3}
\end{figure}

\clearpage

\begin{table*}[!hbp]
\caption{The predicted carrier mobilities of P/h-BN heterostructure
, phosphorene and P/2h-BN heterostructure. Types ``e" and ``h"
represent the ``electron" and ``hole'', respectively. $m^{*}_{x/y}$
(in unit of free electron mass $m_{0}$) is the effective mass
along the  $\Gamma$-X/$\Gamma$-Y direction.
$C_{x/y-2D}$ is the 2D elastic modulus for the
 $\Gamma$-X/$\Gamma$-Y direction which is in unit of Jm$^{-2}$. $E_{ix/y}$ (in unit of eV)
is the deformation potential at the $\Gamma$ point along the $\Gamma$-X/$\Gamma$-Y direction.
$\mu_{x/y-2D}$ (in unit of 10$^{3}$ cm$^{2}$V$^{-1}$s$^{-1}$) is the carrier mobility
along the  $\Gamma$-X/$\Gamma$-Y direction at room
temperature. It is also noticed that the carrier
mobilities of phosphorene are calculated in a supercell as same as that used in
P/h-BN. And the results are similar to those calculated in a unit cell.}

\begin{tabular}{ccccccccccc}
\hline
\hline
Material & type & $m^{*}_{x}$  & $m^{*}_{y}$   & $C_{x-2D}$ & $C_{y-2D}$ & $E_{ix}$ & $E_{iy}$ & $\mu_{x-2D}$ &$\mu_{y-2D}$ \\
\hline
 P/h-BN & e     &       0.21 &  0.43 &   321.41  & 369.86  & 2.21 & 6.58 & 22.34  &  1.42\\
        & h      & 0.18 & 2.99 &  321.41 & 369.86 & 2.83 & 0.24 & 6.53  &  59.96     \\
phosphorene & e & 0.16 & 1.31  & 30.24 & 96.16 & 1.49 & 6.01 & 4.00 & 0.09\\
    & h & 0.15 & 3.28  & 30.24 & 96.16  & 2.80 & 0.58 & 0.79 & 2.67 \\
P/2h-BN &e     &0.20   &0.67   &555.52   &648.00 &  2.09& 6.71& 37.32 &1.26 \\
        & h     &0.18   &3.74   &555.52  &648.00 &  2.71 &0.71 & 11.00&9.00 \\
2P/h-BN &e       &0.22&0.31    &357.29  &466.41 &3.68 &  6.64& 8.57& 2.44\\
        &h         &0.18  &1.78 &  357.29 &466.41 &2.73 &1.37 & 7.80& 4.09\\
\hline
\end{tabular}
\label{tab1}
\end{table*}

\clearpage

\begin{figure}
\includegraphics[width=16cm]{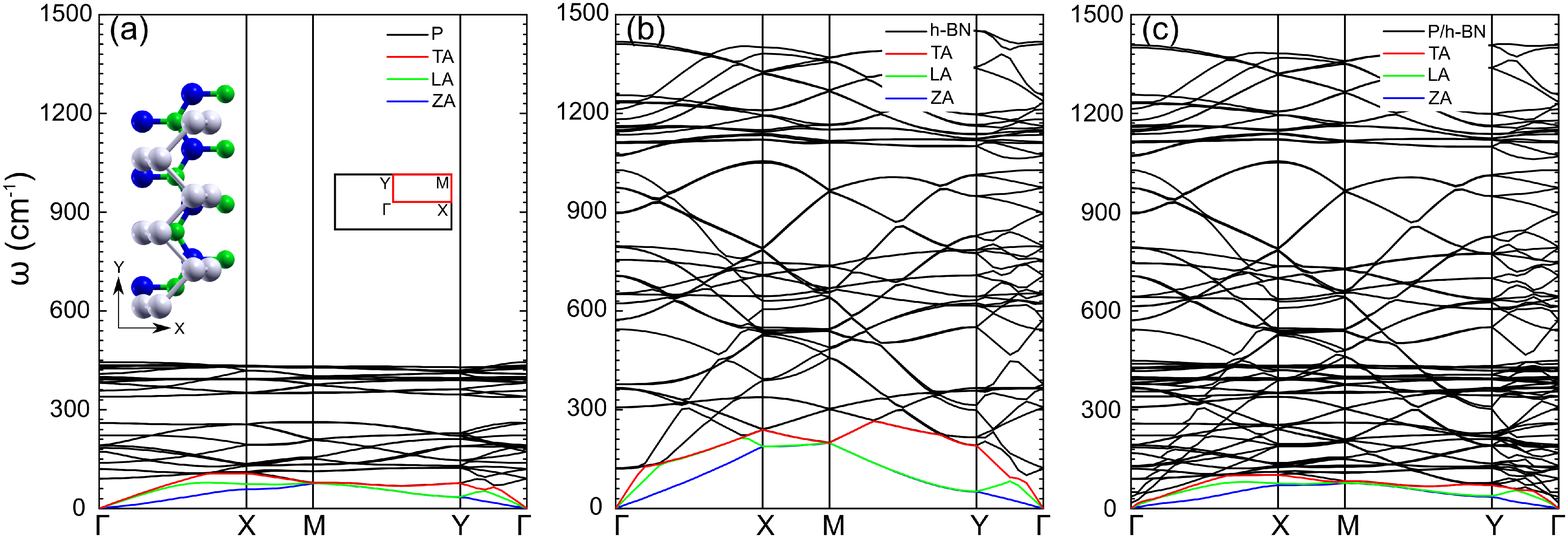}
\caption{(color online). The phonon dispersion of phosphorene (a),
hexagonal boron nitride (b), and P/h-BN heterostructure (c). The inset figure in (a)
is the atomic structure and Brillouin zone of the P/h-BN heterostructure. To be more comparable,
the phonon spectrums are in the same scale.}
\label{fig4}
\end{figure}

\clearpage




\begin{thebibliography}{99}
\bibitem{ling} X. Ling, H. Wang, S. Huang, F. Xia, and M. S Dresselhaus,
{\it PNAS}, {\bf 112}, 4523 (2015); See also references therein.

\bibitem{liu} H. Liu, A. T. Neal, Z. Zhu, Z. Luo, X. Xu, D. Tom\'anek, and P. D Ye,
 {\it \acsnn} {\bf 8}, 4033 (2014).

\bibitem{radi} B. Radisavljevic, A. Radenovic, J. Brivio, V. Giacometti, and A. Kis, {\it \nn} {\bf 6}, 147-150 (2011).

\bibitem{qiao} J. Qiao, X. Kong, Z.-X. Hu, F. Yang, and W. Ji, {\it \nc} {\bf 5}, 4475 (2014).

\bibitem{dasgu} N. Dasgupta, and A. Dasgupta, Semiconductor Devices: Modelling and Technology,
ISBN 81-203-2398-X (2004).

\bibitem{wang} Q. H. Wang, K. Kalantar-Zadeh, A. Kis, J. N. Coleman, and M. S. Strano, {\it \nn} {\bf 7}, 699-712 (2012).

\bibitem{dean} C. R. Dean, A. F. Young, I. Meric, C. Lee, L. Wang, S. Sorgenfrei, K. Watanabe, T. Taniguchi,
P. Kim, K. L. Shepard, and J. Hone, {\it \nn} {\bf 5}, 722-726 (2010).

\bibitem{li}  L. Li, G. J. Ye, V. Tran, R. Fei, G. Chen, H. Wang, J. Wang, K. Watanabe, T. Taniguchi, L. Yang, X. H. Chen, and Y. Zhang, {\it \nn} {\bf 10}, 608-613 (2015).


\bibitem{long} G. Long, S. G. Xu, Z. F. Wu, T. Y. Han, J. X. Z. Lin, J. Y. Shen,
 Y. Han, W. K. Wong, J. Q. Hou, R. Lortz, and N. Wang, arXiv:1510.06518.

\bibitem{li_14}  L. Li, Y. Yu, G. J. Ye, Q. Ge, X. Ou, H. Wu, D. Feng, X. H. Chen, and Y. Zhang, {\it \nn} {\bf 9}, 372-377 (2014).

\bibitem{xia} F. Xia, H. Wang, and Y. Jia, {\it \nc} {\bf 5}, 4458 (2014).



\bibitem{jena}  D. Jena, and A. Konar, {\it \prl} {\bf 98}, 136805 (2007).

\bibitem{pono} L. A. Ponomarenko, R. Yang, T. M. Mohiuddin, M. I. Katsnelson,
 K. S. Novoselov, S. V. Morozov, A. A. Zhukov, F. Schedin, E. W. Hill, and A. K. Geim, {\it \prl} {\bf 102}, 206603 (2009).

\bibitem{konar} A. Konar, T. Fang, and D. Jena, {\it \prb} {\bf 82}, 115452 (2010).

\bibitem{siesta} E. Artacho, E. Anglada, O. Di\'eguez, J. D. Gale, A.
Garc\'ia, J. Junquera, R. M. Martin, P. Ordej\'on, J. M. Pruneda, D. S\'anchez-Portal, and J. M. Soler, {\it \jpcm} {\bf 20}, 064208 (2008).

\bibitem{pbe} J. P. Perdew, K. Burke, and M. Ernzerhof, {\it \prl} {\bf 77}, 3865 (1996).

\bibitem{optvdw}  J. Klimes, D. R. Bowler, and A. Michaelides, {\it \jpcm} {\bf 22}, 022201 (2010).

\bibitem{ca}  D. M. Ceperley, and B. J. Alder, {\it \prl}  {\bf 45}, 566-569 (1980).

\bibitem{df2}  K. Lee, E. D. Murray, L. Kong, B. I. Lundqvist, and D. C. Langreth, {\it \prb} {\bf 82}, 081101 (2010).

\bibitem{c09} V. R. Cooper, {\it \prb} {\bf 81}, 161104 (2010).


\bibitem{isberg} J. Isberg, J. Hammersberg, E. Johansson, T. Wikstr\"om, D. J.
Twitchen, A. J. Whitehead, S. E. Coe, and G. A. Scarsbrook, {\it Science}
{\bf 297}, 1670 (2002).

\bibitem{hwang} E. H. Hwang, and S. Sarma, {\it \prb} {\bf 77}, 115449 (2008).

\bibitem{takagi} S. Takagi, A. Toriumi, M. Iwase, and H. Tango, {\it IEEE Trans. Electron Devices}
{\bf 41}, 2363 (1994).

\bibitem{bruzzone}  S. Bruzzone, and G. Fiori, {\it \apl} {\bf 99}, 222108 (2011).

\bibitem{zhiya} Z. Zhang, J. Xie, D. Yang, Y. Wang, M. Si, and D. Xue, {\it \apex} {\bf 8}, 055201 (2015).

\bibitem{pholimit} Z. Yu, Z. -Y. Ong, Y. Pan, Y. Cui, R. Xin, Y. Shi, B. Wang, Y. Wu, T. Chen, Y. -W. Zhang, G. Zhang, and X. Wang, {\it \am} {\bf 28}, 547-552 (2016)
\bibitem{fei} R. Fei, and L. Yang, {\it \apl} {\bf 105}, 083120 (2014).


\bibitem{cai2} Y. Cai, Q. Ke, G. Zhang, Y. P. Feng, V. B. Shenoy, and Y.-W. Zhang, {\it \afm} {\bf 25}, 2230-2236 (2015)



\bibitem{padilha} J. E. Padilha, A. Fazzio, and A. J. R. da Silva, {\it \prl} {\bf 114}, 066803 (2015).

\bibitem{rivero} P. Rivero, C. M. Horvath, Z. Zhu, J. Guan, D. Tom\'anek, and S. Barraza-Lopez, {\it \prb} {\bf 91}, 115413 (2015).



\bibitem{cai} Y. Cai, G. Zhang, and Y.-W. Zhang, {\it \jpcc} {\bf 119}, 13929 (2015).

\bibitem{cai_sr} Y. Cai, G. Zhang, and Y.-W. Zhang, {\it \sr} {\bf 4}, 6677 (2014).

\bibitem{dai} J. Dai, and X. C. Zeng, {\it \jpcl} {\bf 5}, 1289 (2014).

\bibitem{hsing} C.-R. Hsing, C. Cheng, J.-P. Chou, C.-M. Chang, and C.-M. Wei, {\it \njp} {\bf 16}, 113015 (2014).

\bibitem{shulenburger} L. Shulenburger, A. D. Baczewski, Z. Zhu, J.
  Guan, and D. Tom\'anek, arXiv:1508.04788v1.

\bibitem{rodin} A. S. Rodin, A. Carvalho, and A. H. Castro Neto, {\it \prl} {\bf 112}, 176801 (2014).

\bibitem{herring} C. Herring, {\it Phys. Rev.} {\bf 52}, 365-373 (1937).

\bibitem{hu} T. Hu, and J. Hong, {\it ACS Appl. Mater. Interfaces} {\bf 7}, 23489-23495 (2015).


\bibitem{ashcroft} N. W. Ashcroft, and N. D. Mermin, {\it Solid State Physics}
(Saunders College, Philadelphia, 1976).

\bibitem{pollak} F. H. Pollak, M. Cardona, C. W. Higginbotham, F. Herman, and J. P. Van Dyke,
 {\it \prb} {\bf 2}, 352 (1970).

\bibitem{vdw_hetero} A. K. Geim, and I. V. Grigorieva, {\it Nature} {\bf 499}, 419 (2013).

\bibitem{bachhuber} F. Bachhuber, J. von Appen, R. Dronskowski, P. Schmidt, T. Nilges, A. Pfitzner, and R. Weihrich,
 {\it Angew. Chem. Int. Ed.} {\bf 53}, 11629 (2014).

\bibitem{song} L. Song, L. Ci, H. Lu, P. B. Sorokin, C. Jin, J. Ni, A. G. Kvashnin, D. G. Kvashnin, J. Lou, B. I. Yakobson, and P. M. Ajayan,
{\it \nl} {\bf 10}, 3209 (2010).

\bibitem{lee} C. Lee, X. Wei, W. J. Kysar, and J. Hone, {\it Science} {\bf 321}, 385 (2008).

\bibitem{kudin} K. N. Kudin, G. E. Scuseria, and B. I. Yakobson, {\it \prb} {\bf 64}, 235406 (2001).

\bibitem{bardeen} J. Bardeen, and W. Shockley, {\it \pr} {\bf 80}, 72 (1950).

\bibitem{shockley} W. Shockley, and J. Bardeen, {\it \pr} {\bf 77}, 407 (1950).

\bibitem{nov} K. S. Novosolov, V. I. Fal'ko, L. Colombo, P. R. Gellert,
   M. G. Schwab, and K. Kim, {\it Nature} {\bf 490}, 192-200 (2012).

\bibitem{walu} W. Walukiewicz, H. Ruda, J. Lagowski, and H. Gatos,
  {\it \prb} {\bf 30}, 4571-4582 (1984).
\bibitem{zhu} Z. Zhu, and D. Tom\'anek, {\it \prl} {\bf 112}, 176802 (2014).
\bibitem{jain} A. Jain, and Alan J. H. McGaughey, {\it \sr} {\bf 5}, 08501 (2015).
\bibitem{michel} K. H. Michel, and B. Verberck, {\it \prb} {\bf 83}, 115328 (2011).
\bibitem{jung} S. Jung, M. Park, J. Park, T.-Y. Jeong, H.-J. Kim,
K. Watanabe, T. Taniguchi, D. H. Ha, C. Hwang, and Y.-S. Kim, {\it \sr} {\bf 5}, 16642 (2015)
\bibitem{mcgau} A. J. H. McGaughey, and M. I. Hussein, {\it \prb} {\bf 74}, 104304 (2006).
\end{thebibliography}
\end{document}